\documentclass[aps,twocolumn]{revtex4}
\usepackage{epsfig}
\newcommand{\be}{\begin{equation}}
\newcommand{\ee}{\end{equation}}
\newcommand{\ba}{\begin{array}}
\newcommand{\ea}{\end{array}}

\begin{document}

\title{Simulations of grafted polymers in a good solvent}
\author{Peter Grassberger}
\affiliation{John-von-Neumann Institute for Computing, Forschungszentrum
J\"ulich, D-52425 J\"ulich, Germany}
                                                                                
\date{\today}
\begin{abstract}
We present improved simulations of three-dimensional self avoiding walks
with one end attached to an impenetrable surface on the simple cubic lattice. 
This surface can either be a-thermal, having thus only an entropic effect, 
or attractive. In the latter case we concentrate on the adsorption 
transition, We find clear evidence for the cross-over exponent to be smaller 
than 1/2, in contrast to all previous simulations but in agreement with a 
re-summed field theoretic $\epsilon$-expansion. Since we use the pruned-enriched
Rosenbluth method (PERM) which allows very precise estimates of the partition
sum itself, we also obtain improved estimates for all entropic critical exponents.

\end{abstract}

\maketitle

\section{Introduction}

Single polymers grafted to a plane impenetrable wall have been studied intensely 
for a long time. If the surface is energetically neutral, the main effect 
is a change in the critical exponent governing the scaling of the partition
sum. For an attractive surface the situation is more interesting, as there 
is a second order adsorption transition at a finite strength of the attraction
\cite{eisenriegler82}.

Of particular interest is the cross-over exponent at this transition. Early 
simulations \cite{eisenriegler82,meirovitch88} had given $\phi>0.5$, but it 
was claimed in \cite{hegger94} that this was due to finite size corrections 
to scaling, and that $\phi$ is actually very close to 1/2 (the best estimated
being $\phi=0.496\pm 0.004$). Basically the same conclusion was reached 
recently in \cite{rensburg04}, where an even smaller error bar was obtained, 
$\phi=0.5005\pm 0.0036$, suggesting that $\phi=1/2$ exactly. Since $\phi=1/2$ 
also in $d=2$ and $d\geq 4$, this would mean that $\phi$ is superuniversal,
as it is for branched polymers \cite{janssen95,hsu04}.
But a completely different picture was drawn in 
another recent paper \cite{blumen04}, where it was claimed that $\phi=0.59$. 
To add to the confusion, we should cite field theoretic results. The 
$\epsilon$-expansion with $\epsilon=4-d$ predicts \cite{diehl81,diehl86} 
$\phi = 1/2-\epsilon/16+[16\pi^2-39]\epsilon^2/512 + \ldots$. This would 
give $\phi = 0.68$, if higher order terms were simply omitted, but 
Pad\'e-Borel summation gives $\phi=0.483$ \cite{diehl98}. On the other hand, 
fixed dimension (massive theory) renormalization group calculations 
\cite{diehl94,diehl98} give $\phi = 0.52$. All these calculations have 
errors which are hard to pin down, and the authors of \cite{diehl98} preferred
the value 0.52 over 0.483. Anyhow, these renormalization group calculations 
strongly suggest that $\phi$ is not superuniversal.

It is the purpose of the present work to clarify the situation by means of 
much more precise simulations. We will find that indeed $\phi$ is definitely 
smaller than 1/2 (as predicted by the resummed $\epsilon$-expansion, but not 
by the supposedly more reliable massive field theory approach). In addition, 
we shall also provide precise estimates
for the location of the adsorption transition and for the entropic exponents. 
The latter will be done both for thermal surfaces (at the adsorption transition 
point) and for a-thermal surfaces.
 
We model the polymer by a self avoiding walk (SAW) of $N-1$ steps on a simple
cubic lattice with restriction $z\geq 0$. There is an energy $-\epsilon$ for 
each monomer (site) at $z=0$, the first monomer is located at ${\bf x}=0$.
The temperature is taken to be $T=1$, so that the Boltzmann factor for each 
contact with the surface is $q = e^\epsilon$. The adsorption transition
is at $q=q_c > 1$. For the simulations we use the pruned-enriched Rosenbluth
method (PERM) \cite{g97,permreview}. This is a recursively (depth-first) 
implemented sequential sampling algorithm with re-sampling \cite{liu}. It is 
similar to the algorithm used in \cite{hegger94}, but it is faster and much 
easier to use. Its main advantage over conventional Markov chain Monte Carlo (MC)
methods is that it gives very precise estimates of the partition sum, without
any need for thermodynamic integration or the like. To minimize statistical
errors and speed up the algorithm, we use Markovian anticipation 
\cite{Frauenkron98,hsu03}. We simulated $\approx 6.6\times 10^8$ walks with
$N=8000$ for $q=1$ and $\approx 9.5\times 10^8$ walks at $q\approx q_c$. In
both cases, about 1.6 \% of these walks were strictly independent. 
Altogether this needed $\approx 5000$ hours CPU time on fast (3GHz) PCs. 
During the runs with $q>1$, results at slightly different values
of $q$ were obtained by re-weighing on the fly, so that one run made with
$q\approx q_c$ gave results at three close values of $q$. The critical
point $q_c$ was then found by interpolation.

The partition sum is written as $Z_1(N,q) = \sum_m C_{Nm} q^m$, where 
$C_{Nm}$ is the number of configurations with $m$ contacts with the wall,
$z_{i_k}=0$ for $k=1,\ldots m$, and the subscript `1' indicates that one 
end is grafted. For $q<q_c$ it scales as
\be
   Z_1(N,q) \sim \mu^N N^{\gamma_1-1}                         \label{Z_1}
\ee
with $\mu$ and $\gamma_1$ independent of $q$, but with a $q$-dependent 
prefactor.

Near the adsorption transition, $Z_1(N,q)$ should scale as \cite{eisenriegler82}
\be
   Z_1(N,q) \sim \mu^N N^{\gamma^s_1-1} \Psi[(q-q_c)N^\phi]. \label{Z_c}
\ee
where $\Psi(z)$ is analytic for finite $z$ and $\lim_{z\to -\infty} \Psi(z)$
is finite and positive. Notice that the dominant exponential growth of 
$Z_1(N,q)$ with $N$ is the same as for ordinary SAWs, as long as $q\leq q_c$.
Taking the derivative of $\ln Z_1(N,q)$ with respect to $q$ and 
setting $q=q_c$ thereafter, we obtain for the average energy exactly 
at the critical point
\be
   E_N(q_c) = \langle \epsilon m \rangle \sim N^\phi.  \label{E}
\ee
Taking two derivatives we would obtain a scaling ansatz for the specific
heat which is often used to estimate $\phi$ and other aspects of the critical
behavior. We will {\it not} use it in the present paper, since 
Eqs.(\ref{Z_c}) and (\ref{E}) give much more precise results, as already
found in \cite{hegger94,hsu04}. We refer to \cite{hsu04} (which deals with
the analogous problem for branched polymers) for a detailed
discussion. Notice that Eq.(\ref{Z_c}) cannot be used
with Markov chain MC methods, since the latter do not give simple and 
precise estimates of $Z_1(N,q)$ itself.

\section{Results: A-thermal Walls}

From previous simulations \cite{hegger94,sutter97} we know that $\mu \approx
4.68404$, but the present simulations have higher statistics, therefore we should
first estimate the critical exponent $\gamma_1$ in such a way that neither $\mu$
nor the unknown prefactor in Eq.(\ref{Z_1}) affect the value. For this we 
form the triple ratios \cite{sutter97}
\be
   \gamma_{1,{\rm eff}}(N) =
       1+{4\ln Z_1(N)-3\ln Z_1(N/3)-\ln Z_1(3N) \over \ln 9}  \label{gamma1}
\ee
(with $Z_1(N)\equiv Z_1(N,1)$), which should tend to $\gamma_1$ as $N\to\infty$.
The leading corrections to Eq.(\ref{Z_1}) (and thus also to Eq.(\ref{gamma1}))
should scale as $1/N^\Delta$ with the same exponent $\Delta\approx 0.5$ holding 
also for SAWs in absence of a surface \cite{li95,belohorec97,hsu03}. This would 
suggest that we should get a straight line when plotting 
$\gamma_1(N)$ against $1/\sqrt{N}$. Unfortunately 
this is not true, due to the presence of very large analytic corrections 
$\propto 1/N, 1/N^2, \ldots$. As seen from Fig.~1, they shift the effective
correction to scaling exponent to $\Delta_{\rm eff}\approx 0.7$, and the 
extrapolation to $N\to\infty$ gives
\be
   \gamma_1 = 0.6786 \pm 0.0012
\ee
This is in good agreement with the best previous MC estimate $0.679\pm 0.002$ 
\cite{hegger94} and with the field theoretic result $0.680$ \cite{diehl98}.

%Fig. 1
\begin{figure}
  \begin{center}
   \psfig{file=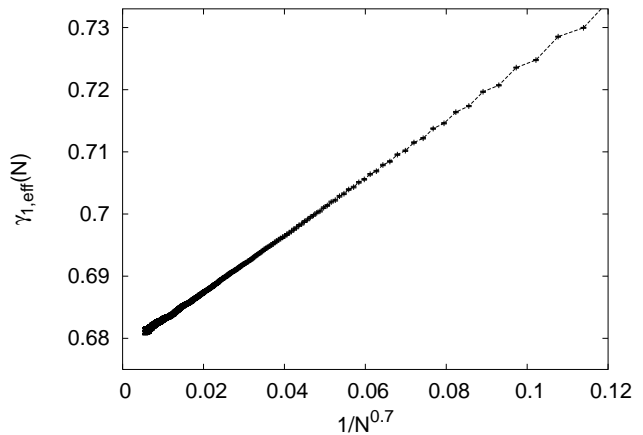,width=6.cm,angle=270}
   \caption{Effective exponents defined by Eq.(\ref{gamma1}), plotted against 
    $1/N^{0.7}$.}
\label{fig-gamma1}
\end{center}
\end{figure}

%Fig. 2
\begin{figure}
  \begin{center}
   \psfig{file=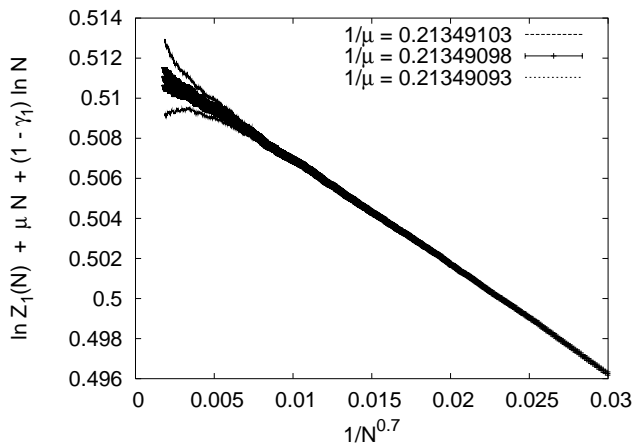,width=6.cm,angle=270}
   \caption{The combination $\ln Z_1(N) - N\ln \mu +(1-\gamma_1)\ln N$, with 
    $\gamma_1$ determined from Fig.~1 and for three candidate values of $\mu$,
    plotted against $1/N^{0.7}$. Error bars are plotted only for the central
    curve, but errors are the same (within the precision of drawing) for all 
    three curves.}
\label{fig-Z1}
\end{center}
\end{figure}

Using this value of $\gamma_1$, we next show in Fig.~2 the quantity 
$\ln Z_1(N) -aN +(1-\gamma_1)\ln N$ for different values of $a$, 
plotted again versus $1/N^\Delta$. Again these data should extrapolate to 
$1/N\to 0$ with a finite slope, if $a=\ln \mu$. From Fig.~2 we see that 
\be
   1/\mu = 0.21349098 \pm 0.00000005 \quad (\mu =4.6840386 \pm 0.0000011),
\ee
where the error includes the uncertainty in the estimate of $\gamma_1$. This 
is the most precise value of the critical fugacity of SAWs on the simple 
cubic lattice published up to now. It is about 80 times more precise than 
the best estimate from exact enumerations \cite{macdonald00}.

In addition to the partition sum for SAWs grafted at one end to the surface,
we can also study the partition sum $Z_{11}(N)$ where both ends are constrained
to have $z=0$. In analogy to Eq.(\ref{Z_1}), it should scale as 
\be
   Z_{11}(N,q=1) \sim \mu^N N^{\gamma_{11}-1}\;.
\ee
The exponent $\gamma_{11}$ should be related to previous exponents by the Barber
relation \cite{barber73}
\be
   \gamma -2 \gamma_1 + \gamma_{11} +\nu =0
\ee
where $\gamma$ and $\nu$ are the entropy and Flory exponents ($R\sim N^\nu$)
for ungrafted SAWs in the bulk. Values of $Z_{11}(N)$ are obtained simply
by summing over those walks for which the $N$-th monomer has $z=0$. Using 
the above estimate of $\mu$ and different candidate values of $\gamma_{11}$, 
we plot in Fig.~3 the analogous quantity to that shown in Fig.~2 for singly
grafted chains. We find
\be
   \gamma_{11} = -0.390 \pm 0.002\;,
\ee
to be compared to the previous MC estimate $-0.383\pm 0.005$ \cite{hegger94} 
and to the field theoretic prediction $-0.388$ \cite{diehl98}.
Using the most precise previous estimates of $\gamma$ and $\nu$ from 
\cite{star03} ($\gamma=1.1573\pm 0.0002$ and $\nu=0.58765\pm 0.00020$), 
we see that the Barber relation is indeed perfectly satisfied,
\be
   \gamma -2 \gamma_1 + \gamma_{11} +\nu = -0.0023\pm 0.0031.
%  \gamma -2 \gamma_1 + \gamma_{11} +\nu = 1.1573\pm 0.0002 - 2(0.6786\pm 0.0012)
%      -0.390 \pm 0.002 +0.58765 \pm 0.00020 = -0.00225\pm 0.0031.
\ee

%Fig. 3
\begin{figure}
  \begin{center}
   \psfig{file=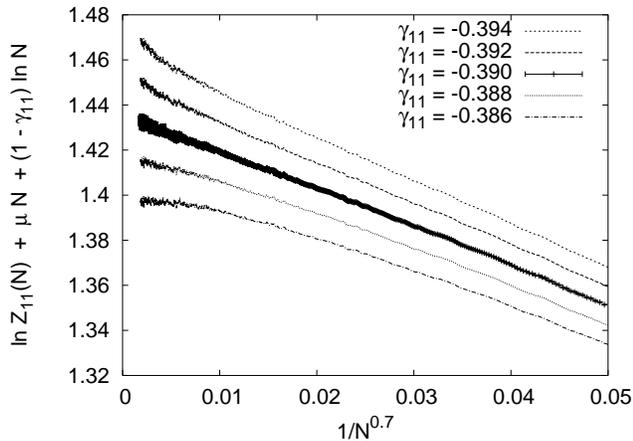,width=6.cm,angle=270}
   \caption{Analogous to Fig.~\ref{fig-Z1}, but for SAWs grafted at both ends
     to the surface, and with the candidate values of $\mu$ replaced by 
     candidate values of $\gamma_{11}$. In order to reduce statistical errors,
     the data are binned with bin width $\Delta N/N \approx 0.01$.}
\label{fig-Z11}
\end{center}
\end{figure}

Root mean square end-to-end distances should scale with the same Flory exponent 
as in the bulk, but with different prefactors and with different corrections 
to scaling. We measured both the components parallel and perpendicular
to the wall for singly grafted SAWs. This time the corrections to scaling 
were $\sim 1/\sqrt{N}$ as expected, i.e. there are much smaller analytic 
corrections. Results are shown in Fig.~4, where we divided averaged square 
distances by $N^{2\nu}$ and plotted them against $1/\sqrt{N}$. The ratio 
$\langle z_N^2 \rangle / \langle x_N^2+y_N^2\rangle$ increases with $N$ as 
found also in \cite{odenheimer04}, but it converges for $N\to \infty$ to a 
finite value, $0.938\pm 0.002$. In contrast to \cite{odenheimer04} we see 
no indication that either $\langle z_N^2\rangle$ or $\langle x_N^2+y_N^2\rangle$
scales with an exponent different from the bulk Flory exponent.

%Fig. 4
\begin{figure}
  \begin{center}
   \psfig{file=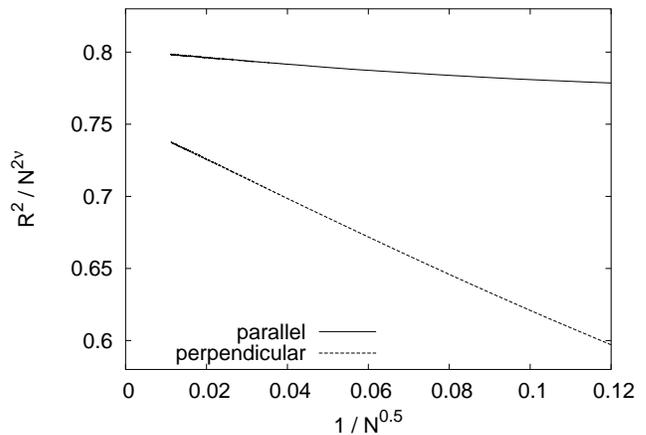,width=6.cm,angle=270}
   \caption{$\langle x_N^2+y_N^2\rangle / N^{2\nu}$ (upper curve) and 
     $\langle z_N^2\rangle / N^{2\nu}$ (lower curve) against $1/\sqrt{N}$.
     Error bars are smaller than the thickness of the lines.}
\label{fig-R1}
\end{center}
\end{figure}

\section{Results: Attractive Walls}

According to Eq.(\ref{Z_c}), the exponential growth of the partition sum
with $N$ is the same at $q_c$ as at $q=1$, i.e. we do not need the triple 
ratio Eq.(\ref{gamma1}) to estimate $\gamma_1^s(N)$. Instead we can use 
\be
   \gamma_{1,{\rm eff}}^s(N,q) = 1 + {\ln[Z_1(2N,q)/Z_1(N/2,q)/\mu^{3N/2}] 
     \over \ln 4}                                            \label{g-c}
\ee
where the ratio between the two partition sums eliminates the unknown prefactor. 
The critical point is characterized by the fact that 
$\gamma_{1,{\rm eff}}^s(N,q)$ diverges for $q>q_c$ when $N\to\infty$, 
converges slowly to $\gamma_1$ for $q<q_c$, and converges to a constant value 
larger than $\gamma_1$ exactly at $q_c$. Values of $\gamma_{1,{\rm eff}}^s(N,q)$
obtained by means of Eq.(\ref{g-c}) are shown in Fig.~5. As in the next 
figures to follow, we plotted it against $1/\sqrt{N}$ since there was no 
different unique value of $\Delta$ which fitted all observables, and $\Delta=1/2$
was not worse overall than other values.

%Fig. 5
\begin{figure}
  \begin{center}
   \psfig{file=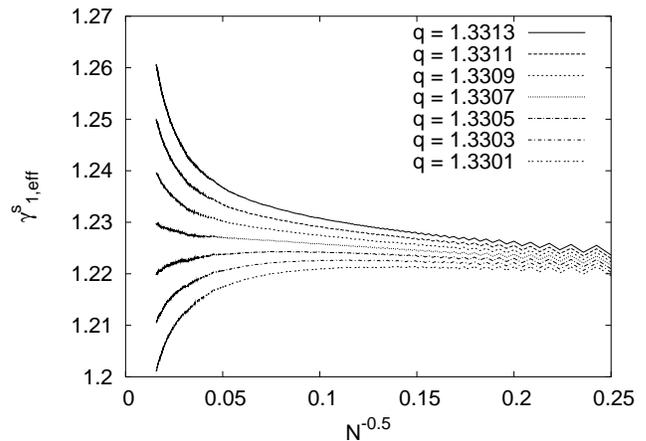,width=6.cm,angle=270}
   \caption{Effective exponents $\gamma_{1,{\rm eff}}^s(N,q)$, plotted versus
     $1/\sqrt{N}$, for several values of $q$.}
\label{gamma-c}
\end{center}
\end{figure}

From Fig.~5 we see that $q_c \approx 1.3307$ and $\gamma_1^s\approx 1.23$, in 
good agreement with the estimates of \cite{hegger94}. More precise values will 
result by combining also the information from other observables.

%Fig. 6
\begin{figure}
  \begin{center}
   \psfig{file=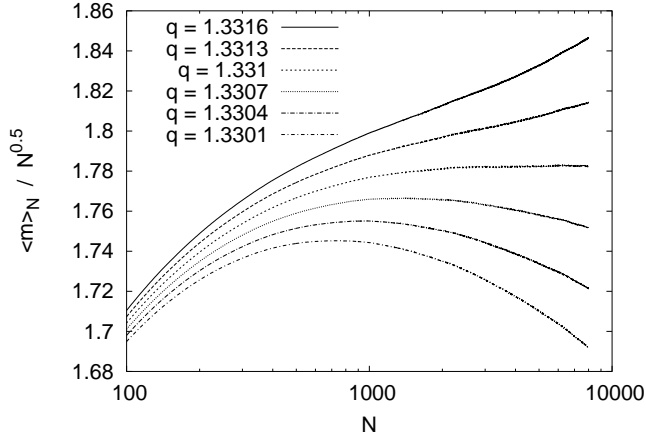,width=6.cm,angle=270}
   \caption{Average number of contacts with the wall, divided by $\sqrt{N}$.
     This would suggest $\phi\approx 1/2$, if we would assume that 
     corrections to scaling are negligible for $N>4000$.}
\label{m-c}
\end{center}
\end{figure}

%Fig. 7
\begin{figure}
  \begin{center}
   \psfig{file=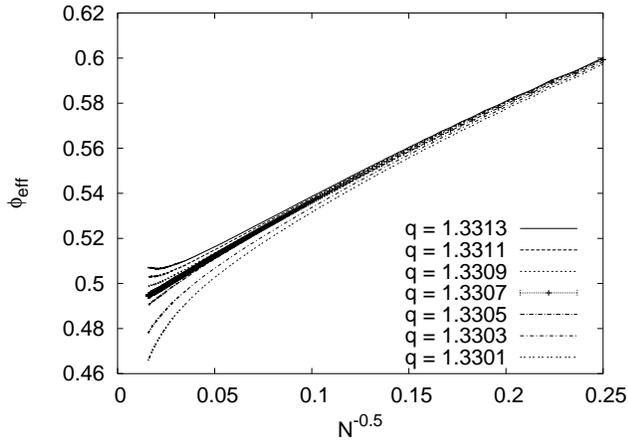,width=6.cm,angle=270}
   \caption{Effective cross-over exponent as defined in the text, plotted
     versus $1/\sqrt{N}$.}
\label{phi-c}
\end{center}
\end{figure}

The next quantity we looked at is the average energy, or rather the average 
number of sites in contact with the wall. Plotting simply $\langle m \rangle 
/ N^\phi$ would suggest $\phi=1/2$ and $q_c=1.331$, see Fig.~6. But this does 
not take into account the fact that corrections to scaling are very large and 
should still be important even for the largest $N$. Indeed, when defining an 
effective exponent by $\phi_{\rm eff}(N,q) = (\ln 4)^{-1} \ln[E_{2N}(q)/E_{N/2}(q)]$
and plotting it against $1/\sqrt{N}$ (Fig.~7), we see that it extrapolates
clearly to a value $< 1/2$, and that $q_c$ is closer to the value $1.3307$ 
found from Fig.5.

%Fig. 8
\begin{figure}
  \begin{center}
   \psfig{file=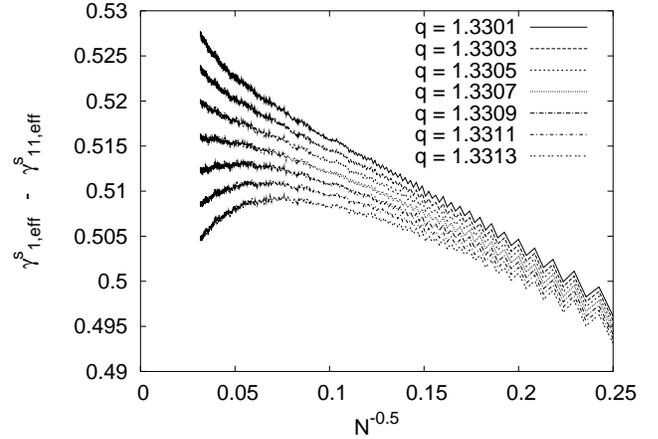,width=6.cm,angle=270}
   \caption{Difference between the effective entropic exponent for polymers
     grafted at one end, and the exponent for polymers grafted at both ends,
     again plotted versus $1/\sqrt{N}$.}
\label{dgam-c}
\end{center}
\end{figure}

%Fig. 9
\begin{figure}
  \begin{center}
   \psfig{file=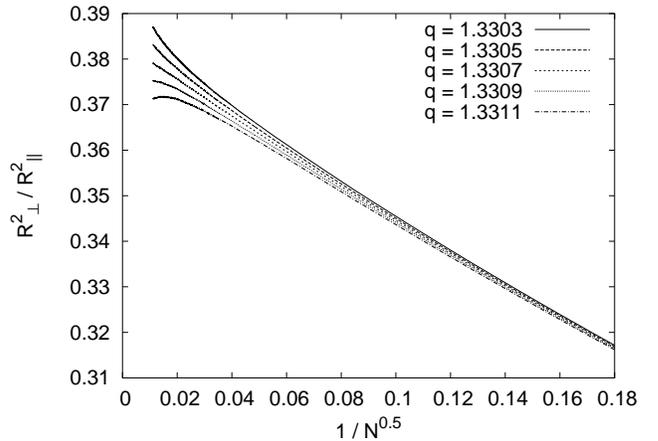,width=6.cm,angle=270}
   \caption{Ratio $\langle z^2\rangle /\langle x^2+y^2\rangle$ between mean
     square end-to-end distances perpendicular and parallel to the wall.}
\label{R-ratio-c}
\end{center}
\end{figure}

Instead of the effective exponent $\gamma^s_{11,{\rm eff}}(N,q)$ defined in 
complete analogy to Eq.(\ref{g-c}), we show in Fig.~8 the difference 
$\gamma^s_{1,{\rm eff}}(N,q)-\gamma^s_{11,{\rm eff}}(N,q)$. This is obtained 
from the ratio between the partition sums for singly and doubly grafted
polymers, and is independent of the precise value of $\mu$.  
Finally, we show in Fig.~9 the ratio between the mean square end-to-end distances 
perpendicular and parallel to the wall. 

Comparing Figs.~5 to 9, we see that all of them (except of course Fig.~6) give the 
same value for $q_c$. Thus we obtain our precise estimate
\be
   q_c = 1.33065\pm 0.00010.
\ee
This is in agreement with the value $1.3310\pm 0.0003$ of \cite{hegger94}, and 
with the recent estimate $1.334 \pm 0.026$ of \cite{rensburg04}. For the anisotropy
of infinitely long chains at the adsorption transition this gives $\langle z^2
\rangle /\langle x^2+y^2\rangle = 0.3845\pm 0.0010$. The best estimates
of the critical exponents are then
\be
   \phi = 0.484 \pm 0.002, \quad \gamma^s_1 = 1.226 \pm 0.002,
\ee
and
\be
  \gamma^s_1 - \gamma^s_{11} = 0.519 \pm 0.003.
\ee
These values are in less agreement with previous estimates. First of all, the 
entropic exponents are slightly smaller (by about 2 standard deviations) than 
the values given in \cite{hegger94}, although these were already lower than all 
previous MC estimates. While $\gamma^s_1$ agrees roughly with the field theoretic 
prediction 1.207 of \cite{diehl98}, $\gamma^s_{11}$ is quite a bit larger 
($0.707$ as opposed 
to $\approx 0.666$). Secondly, and more importantly, the cross-over exponent is 
now clearly less than 1/2, by some 8 standard deviations. The closest previous 
MC estimate was $0.496 \pm 0.004$ \cite{hegger94}, which is off by three sigma. 
The estimate $0.5005 \pm 0.0036$ of \cite{rensburg04} seems clearly excluded 
by our data. It is based on very high statistics of very short ($N\leq 200$) 
chains, which suggests that the corrections to scaling were not taken fully
into account in \cite{rensburg04}.
As mentioned above, $\phi<1/2$ is predicted by the first order term of the 
$\epsilon$-expansion, but not when terms up to $O(\epsilon^2)$ are included and 
not by fixed dimension renormalization group methods. But the resummation 
of the $\epsilon$-expansion done in \cite{diehl98} gave $\phi=0.483$, in 
surprisingly good (and presumably fortuitous) agreement with our result. At 
least, our estimate is comfortably larger than the first order epsilon expansion 
result, $\phi>1/2-\epsilon/16=0.4375$.

\section{Conclusion}

In this paper we have presented Monte Carlo simulations of single grafted 3-d
polymers in a good solvent, both for attractive and for a-thermal walls. 
The sample seems to be the biggest studied so far, both as concerns
the number of chains simulated and their lengths. By using PERM which gives
precise estimates of the partition sum, we could use the partition sum itself
(instead of the specific heat) to locate the critical adsorption point and the 
critical exponents. Our estimates continue the decrease with increased 
statistics observed already in \cite{hegger94}. Our most interesting result
is that the cross-over exponent is clearly less than 1/2, in contrast to all 
previous simulations and to the best estimates from field theory.

I am indebted to Hsiao-Ping Hsu and Walter Nadler for discussions and for
carefully reading the manuscript. I also thank Dieter W. Heermann for sending 
me Ref.~\cite{odenheimer04} prior to publication.

\end{document}